# Transfer Printing of CVD Graphene FETs on Patterned Substrates

T. S. Abhilash[1], R. De Alba[1], N. Zhelev[1], H. G. Craighead[2] and J. M. Parpia[1,a]

[1]Department of Physics, Cornell University, Ithaca, New York 14853, USA.
[2]School of Applied and Engineering Physics, Cornell University, Ithaca, New York 14853, USA.

We describe a simple and scalable method for the transfer of CVD graphene for the fabrication of field effect transistors. This is a dry process that uses a modified RCA-cleaning step to improve the surface quality. In contrast to conventional fabrication routes where lithographic steps are performed after the transfer, here graphene is transferred to a pre-patterned substrate. The resulting FET devices display nearly zero Dirac voltage, and the contact resistance between the graphene and metal contacts is on the order of $910 \pm 340$ $\Omega$ μm. This approach enables formation of conducting graphene channel lengths up to one millimeter. The resist-free transfer process provides a clean graphene surface that is promising for use in high sensitivity graphene FET biosensors.

___________________________________

[a] jmp9@cornell.edu

## I. INTRODUCTION

Graphene, a single planar sheet of carbon atoms has attracted great attention due to its remarkable electrical properties[1]. Large area graphene channel field effect transistors (FET) with pristine surfaces (as characterized by a Dirac voltage close to zero gate-bias), with low noise, and operated in liquid-gating mode, are attractive for biosensing applications[2-7]. Graphene grown through chemical vapor deposition (CVD)[8, 9] can be employed for scaling up graphene channel devices to arbitrarily large dimensions. By scaling up the sensor size, the baseline noise level can be reduced[5, 10] and the devices can be easily adapted for liquid-gated bio sensing applications. Methodologies currently used to transfer CVD graphene require wet chemical etching of the metal substrate on which the graphene is grown and involve the use of polymethylmethacrylate (PMMA) as a support layer to facilitate transfer to the target substrate. This process is problematic for two reasons: 1) The etching process can produce residues that become trapped at the graphene-substrate interface. 2) The use of PMMA introduces contaminates on the top-side graphene surface, degrading the electrical performance of the transistors. Moreover, to fabricate devices in the conventional process, graphene is transferred first followed by the patterning of electrical contacts. Patterning exposes the graphene to polymers that necessitate additional cleaning steps to achieve desirable electrical properties such as a low gate voltage at which the minimum conductance ("Dirac peak") is manifested. A high quality transfer of CVD graphene for FETs has been demonstrated using a modified-Radio Corporation of America (RCA) cleaning method that removes contaminations from the conventional wet chemical etching transfer method[11]. Though this approach provides a clean graphene-substrate interface, it requires the use of a polymer layer to achieve both the transfer of the graphene and the subsequent patterning of electrical contacts. Imperfect removal of the polymer from the top-side of the graphene often results in a residue that compromises the graphene surface quality and the electrical characteristics of the resulting transistor-device. Annealing at high temperature in a $H_2$/Ar atmosphere can remove most of the polymer[12-14], but the trace residues result in variable and often large values of the gate voltage at which the Dirac peak is observed. These characteristics limit the sensitivity of graphene-based sensors[2-7]. Thus it was seen to be desirable when devising high sensitivity graphene sensors to develop a 'resist-free' approach for both the graphene transfer and post transfer processes.

Dry transfer of graphene using polydimethylsiloxane (PDMS)[15-22], thermal release tape[23-25], electrostatic process[26] and pressure sensitive adhesive[27] have also been reported. Although dry transfer printing of graphene can provide a nearly contamination-free surface, again these approaches require post-transfer patterning and exposure to resists. In this paper we report a reliable method for CVD graphene transfer, using a combination of dry transfer by PDMS, and a modified-RCA-cleaning approach. In contrast to the previous reports, we employ PDMS assisted graphene transfer to pre-patterned source-drain electrodes on Si/SiO$_2$ wafers, which circumvents contact between lithographic resists and graphene. This method is scalable and graphene channel FETs with channel lengths as large as one millimeter can be fabricated. Transistor transfer characteristics (source–drain conductance versus back gate voltage) exhibit a clear Dirac peak close to zero back gate voltage. Morphology analysis of graphene after transfer, using Atomic Force Microscopy (AFM), Raman spectroscopy and Scanning Electron microscopy (SEM), confirm the presence of a clean graphene monolayer transferred onto the substrate. The contact resistance ($R_c$) between graphene and array of Ti/Au metal contacts is on the order of 910 ± 340 Ω μm.

## II. EXPERIMENTAL

Our graphene transfer process is schematically depicted in Fig. 1. It begins with the growth of graphene films on Cu foils using CVD[8, 9] and the films obtained were characterized by SEM and Raman spectroscopy. The carbon deposition on the back-side of copper foil was removed by etching using oxygen plasma. Self-prepared PDMS as well as commercially available PDMS (thickness ~50 μm) material (Gelfilm from Gelpak) were used for the graphene transfer with equal success. The Cu/graphene stack was placed on a PDMS block, with the graphene face in contact with the PDMS. The copper foil was gently pressed using a Teflon roller to adhere the graphene face to the PDMS.

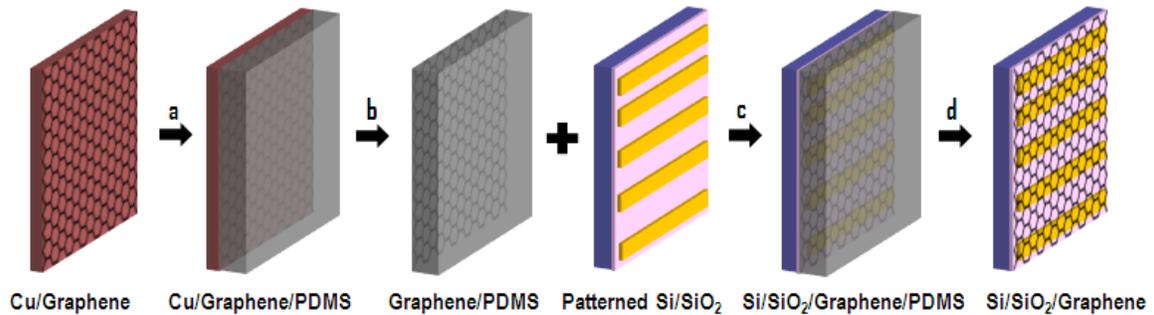

Fig.1: Schematic illustration of CVD graphene transfer process: a) Adhesion of Cu/graphene to PDMS, b) Etching of the Cu, c) Attachment of graphene/PDMS to Si/SiO$_2$, d) PDMS removal. After the copper is etched, a modified-RCA process was carried out to clean the graphene-Cu interface, and then graphene films were transferred onto a source-drain patterned Si/SiO$_2$ wafer.

After adhesion of the PDMS, the Cu/graphene/PDMS assembly was immersed in Cu etchant (HCl/FeCl$_3$ solution in water) for an hour, to etch the copper foil, followed by a repeated rinsing with deionized (DI) water. The resulting graphene/PDMS assembly was further subjected to a modified-RCA cleaning process to remove etch residues and metal particles[11]. First, the assembly was placed in $HCl/H_2O_2/H_2O$ solution (volume ratio 1: 1: 20) for 15 min, followed by a thorough rinsing with DI-water. Then the graphene/PDMS was transferred to a $NH_4OH/H_2O_2/H_2O$ solution (volume ratio 1:1:20) for 15 minutes, followed by a second rinse with DI-water. Finally, the graphene/PDMS assembly was attached to a target $Si/SiO_2$ substrate with the graphene facing the pre-patterned metal contacts. The whole assembly was heated to $140^oC$ for 10 minutes. The PDMS layer was removed by immersing the $PDMS/graphene/SiO_2/Si$ assembly in methylene chloride. This step was followed by rinsing the $graphene/SiO_2/Si$ with DI-water, and a final blow-drying step to complete the transfer.

The graphene-channel FETs were formed by depositing the graphene onto gold source and drain electrodes, pre-patterned on a p-doped Si wafer. The underlying Si serves as a universal back-gate, with 300nm of thermally grown $SiO_2$ isolating it from the conduction and gate channels. Using standard photo-lithographic techniques, source and drain electrodes were patterned with the distance between source-drain electrodes varied from 50 μm to a few millimeters. Titanium (10 nm) and gold (40 nm) were used for source-drain contact metallization. The device design and circuit diagram used for the electrical characterization are given in section 1 of the supplementary information. Electrical measurements were performed at room temperature under ambient conditions.

Contact resistance values are generally measured using a transfer length measurement (TLM) method[28]. We fabricated a TLM structure with varying channel lengths (5 to 30 μm in steps of 5 μm) on a $Si/SiO_2$ wafer using Ti/Au metal contacts. Graphene was stamped over the array after carrying out the cleaning by following our modified-RCA method described earlier. Then the graphene was etched to obtain a rectangular sample of width (*W*) 50 μm as follows: a thin layer of aluminum (20 nm) was deposited on graphene to avoid direct contact with photoresist material. Then the photoresist was spin coated over the graphene-aluminum composite. Photolithography was used to define the channel region, exposing the resist and aluminum. The exposed resist and aluminum was removed by developer solution and the exposed graphene was etched using oxygen plasma. Finally, the aluminum and resist on top of the graphene layer were removed by flood exposure and subsequent treatment with developer solution. This method did not introduce additional contaminants in the graphene channel.

## RESULTS AND DISCUSSION

The devices were electrically characterized by measuring source-drain current ($I_{SD}$) as function of back gate bias ($V_G$) for fixed source-drain voltage ($V_{SD}$ = 50 mV). The gating curve for a device transferred after the use of a conventional Cu etching method, without the modified-RCA cleaning step is shown in the inset of Fig. 2a. The absence of the Dirac peak is an indication that the graphene layer is highly contaminated. Devices fabricated with the commonly used wet transfer method, where the graphene/polymer stack is scooped up over pre-patterned electrodes after the modified-RCA cleaning steps, also showed a similar response. In contrast, the dry transfer of graphene films to the patterned substrate (and subsequent modified-RCA cleaning) consistently produces devices with low Dirac voltages (Fig. 2a, 2b). This indicates that devices prepared in this fashion are much cleaner than those transferred without modified-RCA cleaning or by the conventional process.

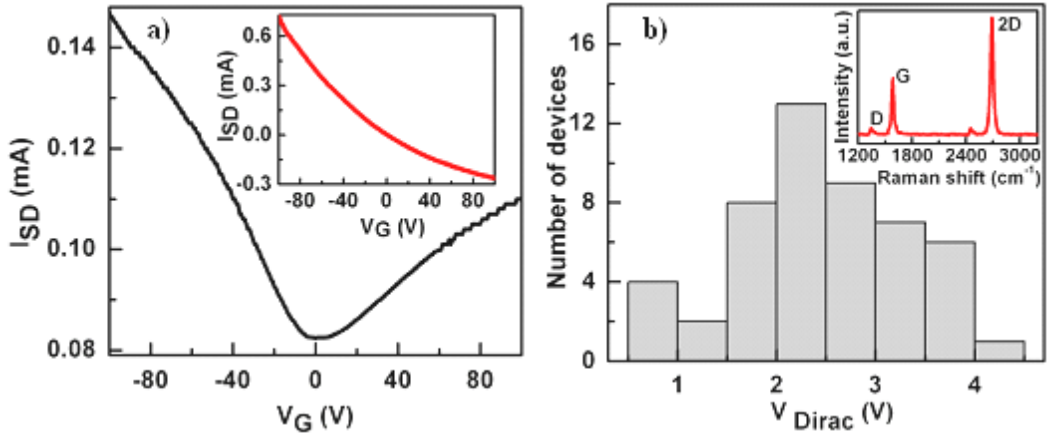

Fig. 2: a) Current-gate voltage ($I_{SD}$−$V_G$) characteristic measurement of a device fabricated with graphene dry transfer and after modified-RCA cleaning. The source-drain spacing of the device used is 50 μm. Inset shows $I_{SD}$−$V_G$ for a device fabricated by conventional Cu etching method. b) Distribution of the voltage of Dirac points of devices fabricated from dry transferred graphene and by using modified-RCA cleaning steps. All devices exhibit their Dirac peak at less than 4.5 volts gate-bias. Inset shows Raman spectrum indicating the transfer of monolayer graphene films to Si/SiO$_2$ substrate.

The reproducibility of the low Dirac voltage was analyzed by measuring 50 devices and the distribution of the observed Dirac voltages is shown in Fig 2b. Dirac voltages were confined to a window of 0.5 to 4.5 V, indicating a narrow distribution. The typical Raman spectrum of a device is shown in Fig. 2b inset. The location and intensity of the characteristic G and 2D peaks signifies the presence of monolayer graphene. The field effect mobility (μ) was extracted using the relationship $\mu = (L/WC_GV_{SD})(\Delta I_{SD}/\Delta V_G)$, where $L$ and $W$ are the graphene channel length (50 μm) and width (50 μm) respectively and $C_G$ is the gate capacitance (11.6 nF/cm$^2$ for 300 nm SiO$_2$). Under ambient conditions, a mobility of 1240 cm$^2$V$^{-1}$s$^{-1}$ was observed. Although the mobility is lower than the best reported mobility of CVD grown graphene[29], it is comparable with the devices fabricated via the modified-RCA cleaning method[11].

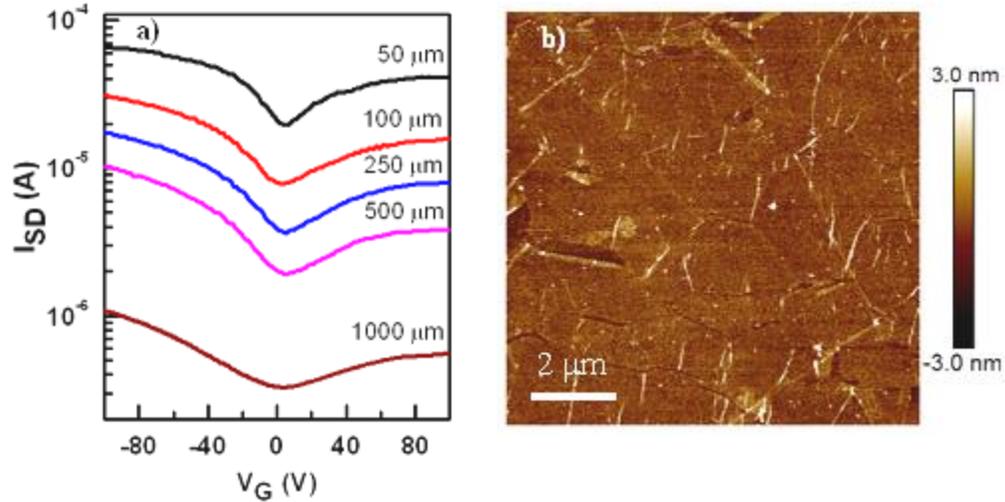

Fig. 3: a) Gating curves of graphene FET array with channel lengths 50–1000 μm. b) AFM image of graphene film transferred on Si/SiO$_2$ substrate.

Though methods are well established for growing graphene over large length scale by CVD[8, 9], most devices have been limited to micrometer lengths[5, 6]. In order to explore the possibility of large-area device fabrication for improved (bio) sensing, where a shift in Dirac peak gate voltage is monitored, FETs were made using the transfer method described above while scaling-up the graphene channel length from 50 μm to several millimeters. Fig. 3a shows $I_{SD}$-$V_G$ measured in a graphene FET device array for different channel lengths. All the devices with a channel length up to 1000 μm show a Dirac peak, and $V_{Dirac}$ is observed at less than 4 V gate-biases across the array. The morphology of the graphene film on the Si/SiO$_2$ substrate was analyzed using AFM and the image obtained (area: 10 μm$^2$) is shown in Fig. 3b. From the image, the surface is found to be nearly flat with a surface roughness of about ± 3 nm. Rips and wrinkles in the transferred graphene are observed, but the graphene sheet is electrically continuous over lengths exceeding one millimeter. The rips and wrinkles may have occurred during Cu/graphene/PDMS assembly preparation step. Surface analysis was done using SEM, and a similar morphology was observed (see section 2 of supplementary information).

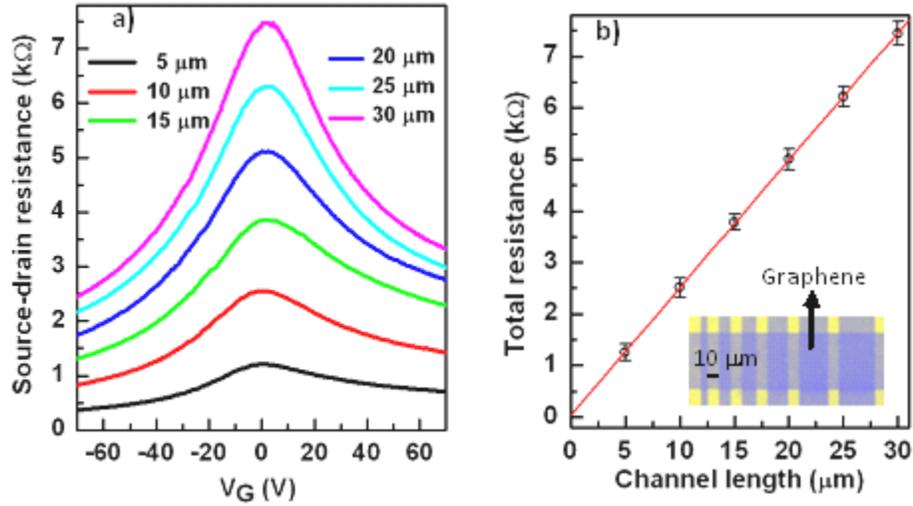

Fig. 4: a) Total resistance between source and drain electrodes in a graphene FET array as a function of back gate bias. Dirac voltage occurs at a gate bias of $3 \pm 0.5$ V. b) $R_{total}$ *vs.* channel length changing from 5 – 30 μm, at $V_G = 3$ V. Black circles indicate total resistance and red line represents the linear fit. Inset shows SEM image (false color) of a graphene FET array used for study.

The methodology followed in the majority of reported works (that measure the electrical properties of graphene) involve the transfer of graphene followed by deposition of the metal contacts. In contrast, one of the novelties of our method lies in the transfer of graphene on to pre-patterned metal contacts. The benefit of this approach compared to the reported methods was evaluated by measuring the metal-graphene contact resistance. Fig. 4a represents the $I_{SD}$-$V_G$ measurements with different channel lengths. The total resistance reaches the maximum value at Dirac point, and it is observed at a back gate bias of ~ $3 \pm 0.5$ volts across the array. A plot of source-drain resistance ($R_{total}$) at the Dirac point as a function of varying channel length is shown in Fig. 4b. It showed a linear behavior and the intercept at zero channel length = $2R_c$ ($36.4 \pm 13.6$ Ω) is obtained. The sheet resistance ($R_s$) can be extracted from the slope of a linear fit, and was found to be 12.4 kΩ/□.

There are various reports on contact resistance measurement in graphene channel field effect transistors[30-33]. Depending on many factors such as fabrication schemes, type of metal used, gate bias voltage and measurement conditions, the reported normalized contact resistance ($R_cW$) values vary from 100 Ω μm to few kΩ μm. In comparison to the $R_cW$ values reported, our fabrication method, where graphene is transferred onto a pre-patterned substrate yields a value for $R_cW$ = 910 ± 340 Ω μm for a Ti/Au contact. This value is lower than most reported. Another advantage of the described method is that it does not require post fabrication steps like thermal and current annealing.

## III. CONCLUSION

We have demonstrated a simple, scalable CVD graphene transfer method employing a combination of dry transfer and modified-RCA cleaning methods. In contrast with the conventional fabrication approach, graphene was transferred to a pre-patterned substrate. FET devices exhibited Dirac voltage close to zero gate bias, and the contact resistance between the graphene and metal was measured as 910 ± 340 Ω μm. This approach enables the FET channel length to be scaled up to devices of length one millimeter. Also, the absence of any resist layers during fabrication steps guarantees a cleaner graphene surface as characterized by a low $V_G$ at which the Dirac peak occurs. Chemical and bio-sensors which rely on the shift of the $V_G$ associated with the Dirac peak would benefit from the device characteristics obtained by following our improved procedure.

# ACKNOWLEDGMENTS

We acknowledge the support of the staff and use of the facilities of the Cornell NanoScale Science and Technology Facility. T.S.A. would like to thank Vinayakan Ramachandran Nair, AEP, Cornell University for fruitful discussions. We also acknowledge the financial support from the Cornell Center for Materials Research under DMR 1120296 and by the NSF under DMR1202991.